\documentclass[aps, prl, nofootinbib, reprint, showpacs]{revtex4-1}
\usepackage[dvipdfmx]{graphicx}
\usepackage{amsmath,amssymb,color}
\usepackage[vcentermath]{youngtab}
\usepackage{ulem}

\newcommand{\etaW}{\eta_\mathrm{WZ}^{}}

\begin{document}
\title{A scalar hint from the diboson excess?}
\author{Giacomo Cacciapaglia$^{a}$, Aldo Deandrea$^{a,b}$}
 \affiliation{$^{a}$Universit\'e de Lyon, France; Universit\'e Lyon 1, Villeurbanne, France;\\
CNRS/IN2P3, UMR5822, IPNL F-69622 Villeurbanne Cedex, France\\
$^{b}$Institut Universitaire de France, 103 boulevard Saint-Michel, 75005 Paris, France}
    \author{Michio Hashimoto}
 \affiliation{Chubu University, 1200 Matsumoto-cho, Kasugai-shi,  Aichi, 487-8501, Japan}
\date{\today}
\preprint{LYCEN-2015-07}
\begin{abstract}
In view of the recent diboson resonant excesses reported by 
both ATLAS and CMS Collaborations, 
we suggest that a new weak singlet pseudo-scalar particle, $\etaW$, 
may decay into two weak bosons while being produced in gluon fusion 
at the LHC. 
The couplings to the gauge bosons can arise from a Wess-Zumino-Witten 
anomaly term and thus we study an effective model based on the anomaly term
as a well motivated phenomenological model.
In models where the pseudo-scalar arises as a composite state, the coefficients of the anomalous couplings can be related to the fermion components of the underlying dynamics. We provide an example to test the feasibility of the idea.
\end{abstract}
\maketitle

{\it Introduction.---}
Recently, the ATLAS search for resonant diboson ($WW/ZZ/WZ$) 
productions in fully hadronic final states has been reported and 
a discrepancy with the background-only model having $3\sigma$ significance  
is observed around 2~TeV~\cite{Aad:2015owa}.
A similar analysis has been also performed by the CMS
Collaboration~\cite{Khachatryan:2014hpa}, where moderate excesses, 
less than $2\sigma$ significance, 
are found in the same mass range $\sim$ 2TeV. 
In the semi-leptonic decay channel of the diboson resonances, however, 
there seems no excess~\cite{Khachatryan:2014gha,Aad:2015ufa}. 
Also, in the fully leptonic decay channel of the $WZ$ resonance, no significant deviation from the Standard Model (SM) prediction is 
observed~\cite{Aad:2014pha,Khachatryan:2014xja}.
Even if the excesses in the fully hadronic decay channels are not just a fluctuation, there is a possibility that a new resonance around 2 TeV 
contributes 
to some of the channels, while others are populated by misidentification 
of the boson-tagged jet: for instance, one may have a neutral resonance 
that only couples to
the $WW$ and $ZZ$ channels, so that the excess in the other $WZ$ channel is just a contamination owing to uncertainties of the 
tagging selections. 
Indeed the situation is still unclear, but it is worth considering 
the possibility for the explanation of the $3\sigma$ discrepancy 
by the ATLAS, which is statistically most significant, 
with some new physics effects. 

This kind of exploratory exercise allows us to consider different new physics hypotheses and to evaluate the theoretical motivations for 
various models behind such an experimental hint. Several authors suggest that these excesses may be due to the existence of 
a new vector resonance, such as a composite $\rho_T$ or a weakly coupled $Z'$ and $W'$, and/or some other effect~\cite{Fukano:2015hga,Hisano:2015gna,Franzosi:2015zra,
Cheung:2015nha,Dobrescu:2015qna,Aguilar-Saavedra:2015rna,Alves:2015mua,Gao:2015irw,Thamm:2015csa,Allanach:2015hba,Carmona:2015xaa,Brehmer:2015cia,Cao:2015lia,Abe:2015uaa,Dobrescu:2015yba,Chiang:2015lqa}. 
Recently, unitarity bounds for the picture describing the excess with 
new vector resonances were also discussed \cite{Englert:2015oga,Cacciapaglia:2015eea,Abe:2015jra}.
A detailed discussion of the population of all diboson channels via misidentified jets can be found in Ref.~\cite{Allanach:2015hba}.
We want to point out a novel possibility related to the existence of a spin-0 resonance whose couplings match the observed excesses while other couplings are naturally absent, thus giving a well-motivated phenomenological model. The case of a scalar coupling to gauge bosons via higher dimensional operators is discussed in \cite{Chiang:2015lqa}.
We suggest instead that a new weak singlet pseudo-scalar particle $\etaW$, which couples to gauge bosons only via a Lagrangian term inspired by the Wess-Zumino-Witten (WZW) 
anomaly~\cite{Wess:1971yu,Witten:1983tw,Fujiwara:1984mp}, can decay into two weak bosons ($WW/ZZ$ channels) while being produced via gluon fusion at the LHC. This case is theoretically well motivated especially in scenarios of multi-TeV scale strong dynamics, where such states arise as a massive scalar associated to an anomalous global symmetry of the confining dynamics. In the following, we will use it as a guiding line, allowing
the precise values of the couplings to vary in order to explore a larger class of models.
We will rely on the composite scenario to establish to what extent the couplings can be considered ``natural'', in the sense of the order of magnitude in the effective theory when particular assumptions
on the specific underlying model are considered. We assume that this pseudo-scalar $\etaW$ is fermiophobic, i.e., its couplings to the SM 
fermions, in particular to the top quark, are vanishing or tiny. 
In detailed composite models this is a realistic possibility and 
such particles with these properties are 
typically present~\cite{Arbey:2015exa}. 
Moreover, new scalar resonances are typically expected to be lighter than their 
vector counterparts\footnote{It is then easier to evade the S-parameter~\cite{Peskin:1990zt} constraint, because one can push up the scale of a new strong dynamics.}, so it is quite reasonable to see
such a pseudo--scalar particle at a lower mass than new vectors, as in QCD.
It is, however, required to construct an appropriate composite model
beyond an adaptation from existing models, because, for example,
familiar dynamical models do not have 
the $WW$ coupling~\cite{Dimopoulos:1980yf}. 

An obvious question is how to enhance the production cross section of
$\etaW$, which is typically expected to be tiny.
Notice that the color factor $N_c=3$ counts strongly 
in $\pi^0 \to \gamma\gamma$. 
Similarly, such an enhancement factor can 
arise from the degrees of freedom of the constituent fermions of $\etaW$, which emerges via the underlying strong dynamics. It is therefore 
interesting to consider this hypothesis and use data from the diboson excess to have an idea of the coefficients required to explain such an excess,
which in turn can hint at the more fundamental structure beyond the effective model.

We therefore assume that $\etaW$ couples to gluons and the weak bosons, 
as in the case of the anomaly, and does couple not (or only very weakly) to 
the SM quarks and leptons.
Before considering detailed numbers, it is useful to put rough numbers on the model: the excess points to an effective diboson cross section of about $10$ fb. This implies that the production cross section $\sigma (gg \to \etaW)$ should 
be around 100 fb in order to explain the diboson excesses as 
the Branching Ratio to the $WW$ channel is roughly
$\mbox{Br}(\etaW \to W^+ W^-) \sim 2(N_c \alpha_2)^2/(8 \alpha_3^2) \sim 10\%$ when considering an anomaly induced coupling. 
For the $ZZ$ mode, we expect a half of it. Thus the desired situation, $\sigma \cdot \mbox{Br}(\etaW \to W^+ W^-/ZZ)\sim 10$ fb,
can in principle be achieved. 
In this scenario, we regard the $WZ$ excess as a contamination from
the $WW/ZZ$ signals, noticing that zero events in the $WZ$ channel
maximizes a likelihood function in terms of the truth signal
in the $WW/ZZ/WZ$ channels~\cite{Allanach:2015hba}.  
The total width should not be so large, which is constrained 
less than, say, 100~GeV, owing to the one-loop effects
essentially.

One might also worry about the constraint from the diphoton resonance searches, because the pseudo-scalar can also decay into a pair of 
photons, unlike the vector resonance.
The constraints of the diphoton channel for a spin 0 resonance have been studied in the mass ranges from 150 GeV to 850 GeV by 
the CMS Collaboration~\cite{Khachatryan:2015qba}. A similar analysis was also performed by the ATLAS collaboration~\cite{Aad:2014ioa}.
For the high-mass diphoton resonances, 
the CMS Collaborations found the constraint of 
the production cross section times branching ratio 
less than $0.3$~fb for the 2TeV RS graviton~\cite{CMS:2015cwa}. 
The ATLAS Collaborations also performed a similar analysis
and the expected $+2\sigma$ variation limit is $0.5$~fb~\cite{Aad:2015mna}.
Even if we take the same bound for the spin 0 particle, our suggested
explanation is fairly safe against this constraint in any case, because
$\Gamma(\etaW \to \gamma\gamma)/\Gamma(\etaW \to W^+ W^-) 
  \sim \alpha^2/2\alpha_2^2 \simeq 0.03$. 
The decay channel of $\etaW \to Zh$ is potentially dangerous~\cite{Aad:2015yza,Khachatryan:2015bma}, 
if one might expect a similar situation to the two Higgs doublet model,
in which there appears $A \to Zh$ at the tree level.
In our case however, we can safely assume that the mixing
  between the singlet and the Higgs doublet is absent.
The possible constraint from final states with Higgs boson(s) is thus easily avoided.

In the rest of the Letter, we will present a general effective
  description of the model, and a specific dynamical model that may give rise to the desired couplings. Finally we will study in a model independent way the constraint on the couplings, necessary to reproduce the observed diboson excess.

{\it Effective Lagrangian.---}
The action for a weak singlet pseudo-scalar $\etaW$ 
with no hypercharge is
\begin{equation}
  S_\eta = \int d^4x \frac{~1~}{2}
  (\partial_\mu \etaW \partial^\mu \etaW - M_{\eta}^2\, \eta_{\mathrm WZ}^2) + \Gamma_{\mathrm WZW},
\end{equation}
where $M_\eta$ is the mass of the pseudo-scalar singlet and
the WZW term contains the effective Lagrangian for the diboson decay,
$
  \Gamma_{\mathrm WZW} \supset \int d^4x\; {\cal L}_{\eta VV},
$
with\footnote{
A term in the form ${\cal L}_{\eta WB} =
   \kappa^\eta_{WB} \frac{g_2 g_Y}{32\pi^2} \frac{~\etaW~}{F_\eta}
   \epsilon^{\mu\nu\rho\sigma} W_{\mu\nu}^3 B_{\rho\sigma}$
can appear through the EWSB effects, however its coefficient is expected to be suppressed by a $v^2/F_\eta^2$ factor as it violates gauge invariance.}
\begin{eqnarray}
  {\cal L}_{\eta gg} &=&
   \kappa^\eta_g  \frac{g_3^2}{32\pi^2} \frac{~\etaW~}{F_\eta}
   \epsilon^{\mu\nu\rho\sigma} G_{\mu\nu}^a G^a_{\rho\sigma}, 
\label{eqgg}   \\[3mm]
{\cal L}_{\eta WW} &=&
   \kappa^\eta_W \frac{g_2^2}{32\pi^2} \frac{~\etaW~}{F_\eta}
    \epsilon^{\mu\nu\rho\sigma} W_{\mu\nu}^i W^i_{\rho\sigma}, 
\label{eqww} \\[3mm]
  {\cal L}_{\eta BB} &=&
   \kappa^\eta_B \frac{g_Y^2}{32\pi^2} \frac{~\etaW~}{F_\eta}
   \epsilon^{\mu\nu\rho\sigma} B_{\mu\nu} B_{\rho\sigma},
\label{eqbb} 
\end{eqnarray}
where $F_\eta$ denotes the decay constant of $\etaW$, and
the couplings $\kappa^\eta_g$, $ \kappa^\eta_W$ and $\kappa^\eta_B$  
are arbitrary prefactors in the effective description, but they can be calculated in 
specific realizations when the content of the loop terms is
calculated, as we shall see in the next section. 
The couplings $g_3$, $g_2$ and $g_Y$ are, respectively, the gauge coupling 
constants of the strong, weak, and hypercharge groups.

From the previous Lagrangian, the partial widths in the various channels can be easily calculated:
\begin{align}
&\Gamma(\etaW \to gg) = \frac{g_3^4 (\kappa^\eta_g)^2 M_\eta^3}{128 F_{\eta}^2 \pi^5 } 
\label{eqg1} \\
&\Gamma(\etaW \to WW) =
\frac{g_2^4 (\kappa^\eta_W)^2 (M_\eta^2 -4 M_W^2)^{\frac{3}{2}}}{512 F_{\eta}^2 \pi^5 } 
\label{eqg2}\\
&\Gamma(\etaW \to ZZ) =
\frac{g_2^4 c_W^4 (\kappa^\eta_W + \kappa^\eta_B t_W^4 )^2 (M_\eta^2 -4 M_Z^2)^{\frac{3}{2}}}{1024 F_{\eta}^2 \pi^5 } 
\label{eqg3}\\
&\Gamma(\etaW \to Z\gamma) =
\frac{e^2 g_2^2 c_W^2 (\kappa^\eta_W - \kappa_B^\eta t_W^2)^2 (M_\eta^2 - M_Z^2)^{3}}{512 F_{\eta}^2 \pi^5 M_\eta^3} 
\label{eqg4}\\
&\Gamma(\etaW \to \gamma \gamma) =
\frac{e^4  (\kappa^\eta_W+ \kappa_B^\eta)^2 M_\eta^3}{1024 F_{\eta}^2 \pi^5} 
\label{eqg5}
\end{align}
where $c_W \equiv \cos \theta_W$, $t_W \equiv \tan\theta_W$, 
$e=g_2 \sin \theta_W$ with $\theta_W$ being the weak mixing angle, 
and we define $\kappa_\gamma^\eta = \kappa_W^\eta + \kappa_B^\eta$ 
for future reference. 
A naive counting of the coupling constants and of the numerical prefactors immediately shows that the production and decay to gluons will be 
the dominant channel, but the values of the couplings $\kappa^\eta_g$ and $ \kappa^\eta_W$ will play a major role in the phenomenological results.

This effective model allows us to easily calculate the diboson rates at the LHC, and check other constraints on the model. Before showing the numerical results, in the next section we will introduce a simple model of underlying dynamics that may lead to the required phenomenology.

{\it A Vector-like Model.---}
In order to discuss the expected phenomenology, we investigate in more detail the origin of the couplings $\kappa^\eta_g$, $ \kappa^\eta_W$ and $\kappa_B^\eta$. In the following 
we take a simple hypothesis of a vector-like model by giving the factors
counting the fundamental particles in the anomaly loops.
We do not discuss here the origin of the electroweak symmetry breaking,
so that we assume that the SM-like Higgs boson with the mass being
125~GeV emerges as a composite object of the dynamics, or we incorporate it
as an elementary particle.
\begin{table}
 \begin{center}
$$
  \begin{array}{|c||c||ccc|} \hline
    & SU(N) & SU(3)_c & SU(2)_W & U(1)_Y \\ \hline
   Q_L=(Q_1,Q_2)_L & \Yboxdim{6pt} \yng(1,1) & {\bf 3} & {\bf 2} & 0 \\
   Q_R=(Q_1,Q_2)_R & \Yboxdim{6pt} \yng(1,1) & {\bf 3} & {\bf 2} & 0 \\[1mm] \hline
   L_L=(L_1,L_2)_L & \Yboxdim{6pt} \yng(1,1) & {\bf 1} & {\bf 2} & 0 \\
   L_R=(L_1,L_2)_R & \Yboxdim{6pt} \yng(1,1) & {\bf 1} & {\bf 2} & 0 \\[1mm] \hline
   N_L & \Yboxdim{6pt} \yng(1,1) & {\bf 1} & {\bf 1} & 0 \\
   N_R & \Yboxdim{6pt} \yng(1,1) & {\bf 1} & {\bf 1} & 0 \\[1mm] \hline
  \end{array}
$$
  \end{center}
  \caption{Charge assignment for a vector-like model under the new strong dynamics $SU(N)$, and the SM gauge symmetries. The chirality of the fermion field is denoted by a subscript $R$ or $L$.
   \label{charge1}}
\end{table}

Let us study as an example the vector-like model shown in Table~\ref{charge1},
where $SU(N)$ represents a strongly interacting gauge group.
Such a dynamical model with the higher representations of the gauge group
has been studied, for example, in Ref.~\cite{Hong:2004td}.
Of course, we may take the fundamental representation as usual,
if we allow arbitrary large $N$.
We introduce vector-like weak doublets $Q$ and $L$,
with multiplicity $n_Q$ and $n_L$, respectively.
The vector-like fermion $N_{L,R}$ is a weak singlet. 
The total number of flavors is then $N_f = 2N_c n_Q + 2n_L + 1$, where $N_c = 3$ denotes the number of ordinary QCD colors.
A large number of $N_f$ is inappropriate,
because the gauge theory loses asymptotic freedom when the fermion multiplicity is too large.
At the one-loop level, asking for a negative coefficient of 
the $\beta$ function~\cite{Caswell:1974gg}, 
we obtain $N_f < 11N/(4T(R))$, where $T(R)$ is the trace normalization.
For the two-index anti-symmetric representation, $T(R)=(N-2)/2$.
For example, the theory with $N=5$ and $n_Q=n_L=1$ keeps asymptotic freedom. 
A question whether or not such a gauge theory might fall into 
the conformal window is beyond the perturbative approach and would require a dedicated analysis: some indications can be extracted \cite{Dietrich:2006cm}; however, only, a Lattice simulation \cite{DelDebbio:2010zz} can give the final answer.

The Nambu-Goldstone boson $\etaW$ is contained in the $U(1)$ part of
the broken current SU($N_f$)$_L \times$ SU($N_f$)$_R \to$ SU($N_f$)$_V$.
The broken current corresponding to $\etaW$ is
\begin{equation}
  J_5^\mu \sim \bar{Q}\gamma^\mu \gamma_5 Q + \bar{L}\gamma^\mu \gamma_5 L 
  - (N_f-1) \bar{N}\gamma^\mu \gamma_5 N , 
\end{equation}
where we omitted the normalization factor 
of the axial current, which can be absorbed in the definition of $F_\eta$.

We then find
\begin{eqnarray}
  \kappa^\eta_g &=& \frac{1}{2} N (N-1) \cdot 2n_Q ,\\
  \kappa^\eta_W &=& \frac{1}{2} N (N-1) \cdot (N_c n_Q + n_L),
\end{eqnarray}
and $\kappa^\eta_B = \kappa_{WB}^\eta=0$,
where $N_c=3$ denotes the number of color.
For the fundamental representation, 
the factor $N(N-1)/2$ should be replaced by $N$.
The coefficient $\kappa^\eta_\gamma$ of the WZW term 
for the $\eta_W$--$\gamma$-$\gamma$ coupling is 
calculated from the above ones and found as $\kappa^\eta_\gamma=\kappa^\eta_W$
in this specific model.
The number $\kappa^\eta_W/\kappa^\eta_g = 2$ for
$n_Q=n_L=1$ corresponds to the number of the weak doublets over 
that of the quark flavor and will play an important role in 
the diboson excess discussed later.

For the Branching Ratios, we obtain
\begin{equation}
  \frac{\mbox{Br}(\etaW \to W^+ W^-)}{\mbox{Br}(\etaW \to gg)}
  \simeq \frac{2 (\alpha_2 \kappa^\eta_W)^2}{8 (\alpha_3 \kappa^\eta_g)^2}
  \simeq 0.09\,, 
  \label{eqr1}
\end{equation}
for $n_Q=n_L=1$, where we used $\alpha_3 \approx 0.1$ and 
$\alpha_2 \approx 0.03$.
Also,
\begin{equation}
  \frac{\mbox{Br}(\etaW \to \gamma\gamma)}{\mbox{Br}(\etaW \to W^+ W^-)}
  \simeq \frac{(\alpha \kappa^\eta_\gamma)^2}{2 (\alpha_2 \kappa^\eta_W)^2}
  = \frac{\alpha^2}{2\alpha_2^2} \simeq 0.03\,, 
      \label{eqr2}
\end{equation}
due to $\kappa^\eta_\gamma=\kappa^\eta_W$ in this model,
where we used $\alpha=1/128$.
These numbers can be directly compared to the experimental bounds on the diboson excess, and constraints on other channels, most notably dijet and diphoton resonance searches:
\begin{itemize}
\item[-] $\sigma_{gg \to \etaW} \times Br(\etaW \to WW) \sim 10$ fb, from 
the diboson excess at 2 TeV~\cite{Aad:2015owa};
\item[-] $\sigma_{gg \to \etaW} \times Br(\etaW \to \gamma \gamma) < 0.5$ fb, from the searches of a Kaluza--Klein graviton to di-photon (approximate) \cite{Aad:2015mna};
\item[-] $\sigma_{gg \to \etaW} \times Br(\etaW \to gg) < 200$ fb, from the search of dijet resonances (gluons) from a scalar \cite{Khachatryan:2015sja}.
\end{itemize}
Taking ratios of the above bounds, we can extract direct bounds on the ratios of Branching Ratios:
\begin{eqnarray}
 \frac{\mbox{Br}(\etaW \to W^+ W^-)}{\mbox{Br}(\etaW \to gg)} &>&
 \frac{10}{200} = 0.05, \\
 \frac{\mbox{Br}(\etaW \to \gamma\gamma)}{\mbox{Br}(\etaW \to W^+ W^-)} &<&
 \frac{0.5}{10} = 0.05\,,
\end{eqnarray}
which are easily satisfied in this model.

These simplified results clearly show that the fermiophobic pseudo-scalar with the anomalous interactions can explain the diboson excesses without
conflict with the other experimental bounds we discussed. One has to keep in mind, however, that a detailed model built along these lines may require further scrutiny
concerning other bounds, but such a detailed study is worth pursuing only if the present excess will be confirmed by the ongoing LHC run.

{\it Numerical results and discussion.---}
In order to have more detailed numbers we have created 
a FeynRules~\cite{Christensen:2008py,Alloul:2013bka} model and evaluated
the cross sections, branching ratios and decay widths numerically using 
Madgraph~\cite{Alwall:2014hca}. 
Using the following numerical values, $n_Q=1$, $n_L=1$, $N=2$, $N_c=3$, which correspond to 
$ \kappa^\eta_g=2$ and $\kappa^\eta_\gamma = \kappa^\eta_W=4$, and $F_\eta = 1$~TeV,
the production cross section of the $\etaW$ particle is 0.615 fb and its total width 1.12 GeV at LHC with 8 TeV of center of mass energy for a $\etaW$ particle of 
2 TeV of mass.

Using instead $N=5$ and all the other same parameters as in the previous example, increases the couplings by a factor of 10:
$ \kappa^\eta_g=20$ and $\kappa^\eta_\gamma = \kappa^\eta_W=40$, 
while the production cross section and width of the $\etaW$ particle 
are a factor of 100 larger as expected 
(production cross section of 61.5 fb and total width of 112 GeV). 
The results for the branching fractions are given in Table \ref{br2}. 
\begin{table}
 \begin{center}
 $$
  \begin{array}{|c|c|c} \hline
\mathrm{decay\; mode}    & BR  \\ \hline
gg  & 83 \%  \\
WW  & 11.2\% \\
ZZ  & 3.2\% \\
Z\gamma  & 2\% \\
\gamma\gamma  & 0.4\%  \\
\hline
  \end{array}
  $$
  \end{center}
  \caption{Decay modes and branching fraction of the $\etaW$ particle of 2 TeV with $ \kappa^\eta_W/\kappa^\eta_g=2$.
   \label{br2}}
\end{table}
These number are just indications based on a particular choice of
parameters. One can see easily from the previous results that increasing $N$ 
(or decreasing $F_\eta$)
will increase the cross section and allow reaching a value compatible with the excess. 

We consider in the following the parameters $\kappa^\eta_i$ in order to describe and 
bound the model in an effective way without reference to a particular underlying model. 
First, we can impose bounds on the couplings by taking ratios of
Branching Ratios and compare them with the bounds detailed 
in the previous section on the diboson, dijet, 
and diphoton resonant cross sections.
Taking ratios of formulas (\ref{eqg1})--(\ref{eqg5}), we can eliminate the dependency on the cross section, and derive bounds on the couplings 
$\kappa^\eta_i$:
\begin{eqnarray}
 \frac{(\kappa^\eta_W)^2}{(\kappa^\eta_g)^2} &>&
 \frac{1}{5} \frac{~g_3^4~}{g_2^4} \sim 1.45, \\
 \frac{(\kappa^\eta_\gamma)^2}{(\kappa^\eta_W)^2} &<& 
 0.1 \frac{~g_2^4~}{e^4} = \frac{0.1}{\sin^4 \theta_W}\sim 1.86,
\end{eqnarray}
where $g_3 = 1.033$, $g_2 = 0.628$, and $\sin^2 \theta_W = 0.2319$ at an energy of 2 TeV.

\begin{figure}[t!]
\begin{center}
\resizebox{0.47\textwidth}{!}
          {\includegraphics{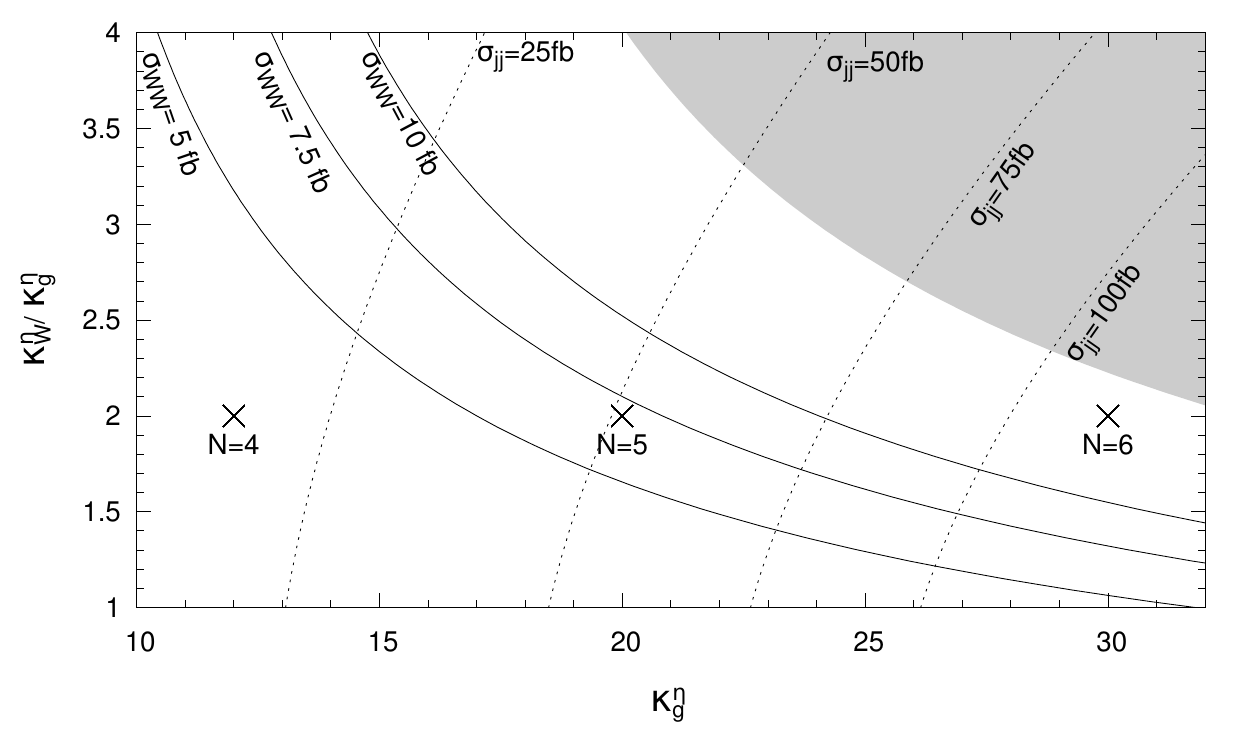}}
\end{center}
\caption{
Cross section times branching ratios on 
the $\kappa^\eta_g$--$\kappa^\eta_W/\kappa^\eta_g$ plane
for $F_\eta = 1$~TeV and $\kappa^\eta_B= 0$.
The shaded region in the right upper area is excluded owing to 
$\sigma (gg \to \etaW) \cdot \mbox{Br}(\etaW \to \gamma\gamma) >$ 0.5fb.
The numbers $N=4,5,6$ represent the corresponding values for 
the vector-like model with $n_Q=n_L=1$.
} \label{500}
\end{figure}

To compute constraints in 
the $\kappa^\eta_g$--$\kappa^\eta_W/\kappa^\eta_g$ plane, 
we need an expression for the cross section:
\begin{equation}
\sigma(gg\to \etaW) = \left(\frac{\kappa^\eta_g}{2}\right)^2 \frac{(1~\mbox{TeV})^2}{F_\eta^2}\; 0.615~\mbox{fb}
\end{equation}
which can be estimated by rescaling our numerical results. 
For the $\mbox{Br}(\etaW \to gg)$ and $\mbox{Br}(\etaW \to WW)$, 
by using Eqs.~(\ref{eqg1})--(\ref{eqg5}) for
$\kappa^\eta_B=0$,
we have
\begin{equation}
\mbox{Br}(\etaW \to gg) \simeq
  \frac{8 g_3^4 (\kappa^\eta_g)^2}
       {8 g_3^4 (\kappa^\eta_g)^2  + 3 g_2^4 (\kappa^\eta_W)^2},
\end{equation}
and
\begin{equation}
\mbox{Br}(\etaW \to WW) \simeq
  \frac{2 g_2^4 (\kappa^\eta_W)^2}
       {8 g_3^4 (\kappa^\eta_g)^2  + 3 g_2^4 (\kappa^\eta_W)^2},
\end{equation}
respectively.
These estimates can change if we introduce
$\kappa^\eta_B, \kappa^\eta_{WB} \ne 0$ in general.
\begin{figure}[t]
\begin{center}
\resizebox{0.47\textwidth}{!}
          {\includegraphics{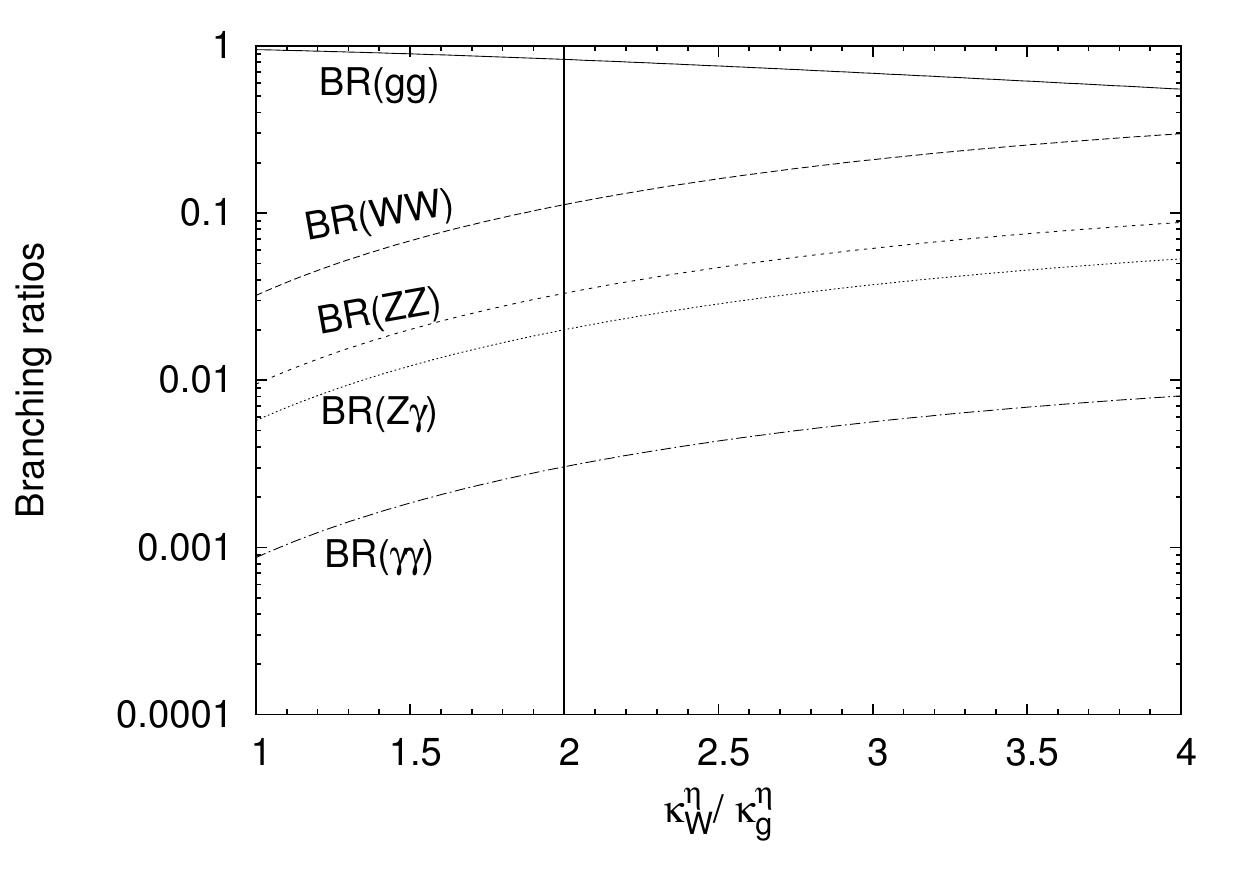}}
\end{center}
\caption{
Branching ratios from the formulae of Eqs.~(\ref{eqg1})--(\ref{eqg5}) for
$\kappa^\eta_B=0$. The vector-like model with $n_Q=n_L=1$ 
corresponds to $\kappa^\eta_W/\kappa^\eta_g=2$. \label{BR}
}
\end{figure}
In Fig.~\ref{500} we show the dijet $\sigma_{jj}$ and 
diboson $\sigma_{WW}$ cross sections for $F_\eta = 1$~TeV
as a function of the $\kappa^\eta_g$ and the ratio $\kappa^\eta_W/\kappa^\eta_g$
(for $\kappa^\eta_B = 0$).
We also show the model predictions for $N=4,5,6$.
In the case of $N=7$, $\sigma_{jj} > $ 200~fb.
The shaded region in Fig.~\ref{500} corresponds to 
$ \sigma(gg\to \etaW) \cdot BR(\etaW \to \gamma\gamma) > $ 0.5fb.
We then find that the model with $N=4$ cannot explain the diboson excess,
on the other hand, the case with $N=5$ does. 
It is fairly safe for $N=6$, although the diphoton production is
slightly large.
The Branching Ratios are depicted in Fig.~\ref{BR}.

In dynamical models, there usually appear many pseudos other than a singlet.
How about the constraint of the color-octet pseudos?
We estimate the difference $\Delta M^2$ of the mass squared by rescaling 
the electromagnetic mass splitting in the $\pi^\pm$--$\pi^0$ 
system~\cite{Farhi:1980xs}, 
\begin{equation}
  \frac{\Delta M^2}{m_{\pi^\pm}^2 - m_{\pi^0}^2} =
  \left(\frac{F_\eta}{f_\pi}\right)^2
  \frac{\alpha_3(F_\eta)}{\alpha}\frac{~3~}{1} ,
\end{equation}
where the factor $3$ is the color Casimir for the octets.
We then find the mass of the color-octet pseudos as 3~TeV with
$F_\eta=1$~TeV, which is consistent with the lower mass bound of
$2.70$~TeV ($2.5$~TeV) by the ATLAS (CMS) Collaborations~\cite{Khachatryan:2015sja}. 

These results show that the possibility of a fermiophobic
pseudo--scalar singlet coupling with anomaly type couplings to the gauge bosons can give an explanation of the 
diboson excess without requiring the artificial suppression of other channels
such as $\etaW \to Zh$. 
The interpretation in terms of a more fundamental model, due to the large 
$\kappa^\eta_g$ and $\kappa^\eta_W$ couplings, requires relatively large representations as, for example, indicated in the vector-like model discussed in 
the previous section. This is not a problem in itself but the detailed model building requires some care in order to avoid other bounds from, for example, electroweak
precision tests or the presence of other states which may be in the
same mass range as the singlet $\etaW$. 
Although there is no $W^3$--$B$ mixing from the model construction,
there may appear large contributions to the trigauge boson couplings
such as $W^+ W^- \gamma$, for example.
The motivation for a further more detailed analysis
will depend on the confirmation or not of the present diboson excess in the near future.

{\it Acknowledgements.---}
We thank the France-Japan Particle Physics Lab (TYL/FJPPL) for partial support. A.D. is partially supported by the ``Institut 
Universitaire de France.''  We also acknowledge partial support from 
the D\'{e}fiInphyNiTi-projet structurant TLF; the Labex-LIO (Lyon
Institute of Origins) under Grant No. ANR-10-LABX-66 and FRAMA 
(FR3127, F\'ed\'eration de Recherche ``Andr\'e Marie Amp\`ere'').

\end{document}